\documentclass[graybox]{svmult}

\usepackage{mathtools}

\usepackage{subfig}

\usepackage{graphicx}
\usepackage{times}
\usepackage{graphicx}
\usepackage{multirow}
\usepackage{breqn}
\usepackage{tabularx}
\usepackage{epstopdf}
\usepackage{mathptmx}       %
\usepackage{helvet}         %
\usepackage{courier}        %
\usepackage{type1cm}        %
\usepackage{makeidx}         %
\usepackage{graphicx}        %
\usepackage{multicol}        %
\usepackage[bottom]{footmisc}%

\authorrunning{A. Adeel, M. Gogate, S. Farooq, C. Ieracitano, K. Dashtipour, H. Larijani, A. Hussain}
\makeindex             %

\begin{document}

\title*{A Survey on the  Role of Wireless Sensor Networks and IoT in Disaster Management}
\author{Ahsan Adeel, Mandar Gogate, Saadullah Farooq, Cosimo Ieracitano, Kia Dashtipour, Hadi Larijani, Amir Hussain}
\institute{Ahsan Adeel, Mandar Gogate, Amir Hussain, Cosimo Ieracitano and Kia Dashtipour are with the Department of Computing Science and Mathematics, Faculty of Natural Sciences, University of Stirling. Hadi Larijani is with the School of Engineering and Built Environment, Glasgow Caledonian University.
\email{ahsan.adeel@stir.ac.uk}
}

\maketitle

\abstract{Extreme events and disasters resulting from climate change or other ecological factors are difficult to predict and manage. Current limitations of state-of-the-art approaches to disaster prediction and management could be addressed by adopting new unorthodox risk assessment and management strategies. The next generation Internet of Things (IoT), Wireless Sensor Networks (WSNs), 5G wireless communication, and big data analytics technologies are the key  enablers for future effective disaster management infrastructures. In this chapter, we commissioned a survey on emerging wireless communication technologies with potential for enhancing disaster prediction, monitoring, and management systems. Challenges, opportunities, and future research trends are highlighted to provide some insight on the potential future work for researchers in this field.}

\section{Introduction}
Albert Einstein once said: “The true sign of intelligence is not knowledge but imagination”. Imagining how technology could be exploited to address the challenging real-world problems is necessary, especially when development of new device or system is considered. Disaster management is one of the challenging real-world problems that sought out emerging technologies. Natural disasters have been visiting every part of the globe and the world has become increasingly vulnerable. Nearly 3 million people worldwide have been killed in past 20 years due to natural disasters, including earthquakes, landslides, floods, cyclones, snow avalanches etc. Intelligent infrastructures to enhance disaster management, community resilience and public safety have become inevitably important, with the aim to save lives, reduce risk and disaster impacts, permitting efficient use of material and social resources, and protect quality of life and economic stability across entire regions. At the leading edge of disaster management initiatives is the collection, integration, management and analysis of an increasingly complex web of multi-modal data and digital information originating from mobile, fixed, and embedded sources. A national program for investment in intelligent infrastructure can achieve dramatic economies of scale and reduce long-term national debt. For example, United Sates spends billions of dollars annually to suppress catastrophic wildfires, which consume millions of acres each year. A single major hurricane or tornado can claim hundreds of lives and cause billions of dollars damage. Intelligent infrastructure technologies such as computer models taught through machine learning and calibrated on big data, including ground-based sensors, streaming video from unmanned aerial vehicles, and satellite imagery, could significantly reduce the social and economic costs of such disasters. Priority should be placed on the areas such as sensing and data collection, communication and coordination, big data modelling frameworks (including analytics and tools for disaster prediction and management), and social computing.

There is a great deal of interest in developing disasters management systems capable of saving lives, properties, and minimise the costly economic investments. In order to develop an effective monitoring infrastructure, the information has to be gathered from different sources. In this context, IoT technology is reporting considerable success \cite{atzori2010internet}. In last few years, innovative real-time monitoring and disasters warning systems are based on the emerging IoT paradigm, in which things (i.e. sensors) are globally interconnected. WSNs as part of IoT have been employed widely for monitoring natural disasters in remote and inaccessible areas \cite{akyildiz2002wireless},\cite{alemdar2010wireless},\cite{chen2013natural}. WSNs use autonomous low-energy sensor nodes capable of measuring and recording surrounding environmental conditions. Each sensor node typically consists of a power supply, a micro-controller, a wireless radio transmission and a set of environmental sensors (i.e. humidity, pressure, temperature). WSNs and IoT together with the recent advances in the Information and Communication Technology (ICT) could develop ever more intelligent and connected infrastructures, where a huge amount of data could be gathered and processed \cite{asimakopoulou2011buildings}. The combination of these heterogeneous resources (gathered from digital infrastructures) and the latest artificial intelligence technology could be used to develop next-generation of disaster management systems.

The rest of the chapter is organised as follows: Section 2 first presents an overview of state-of-the-art WSNs driven disaster monitoring and management systems, including emerging technologies such as 5G, Device to Device communication, Fourth Generation (4G)/LTE, and software defined radio. Section 3 presents an overview of existing IoT standards (LoRa/4G LTE), their limitations, and future research directions. Finally, Section 4 concludes this chapter.

\label{sec:1}

\section{WSN Driven Disaster Monitoring and Management Systems}
\label{sec:2}

\subsection{Applications of Sensor Networks in Disasters Management}
\label{subsec:2}
The application of sensor networks for monitoring natural hazards (such as  floods \cite{shah2011aqua}, wildfires \cite{kung2006drought} or sandstorms \cite{wang2011network}) has become a special research topic for many researchers and engineers. In this context, a lot of work has been focusing on using WSNs for monitoring landslides (e.g. downfalls of a large mass of ground, rock fragments and debris especially in unstable areas where intense rainfalls, floods or earthquakes occur and might cause loss of lives, damage buildings and influence the economy \cite{varnes1984landslide}). Kotta et al. \cite{kotta2011wireless} proposed a WSNs system based on accelerometers for vibrations detection triggered by landslides. Experimental results showed that accelerometers values above 1g (gravity) indicated intense mass sliding and hazardous conditions. Ramesh et al. \cite{ramesh2014design} installed a sensors distributed monitoring system based on 50 geological sensors and 20 wireless sensor nodes to monitor a local zone highly at risk from landslides in India (Idukki, Kerala State). The proposed system was able to provide three level alerts (low, intermediate, high) and its effectiveness was tested during the monsoon season. Terzis et al. \cite{terzis2006slip}  proposed a sensor columns based network to detect the slip surface location and the trigger of a landslide. Lee et al. \cite{lee2017open} presented a slope movement monitoring WSNs system capable of reducing the power consumption during the standby mode (0.05 mA at 3.6 V). Rossi et al. \cite{rosi2011landslide} reported the development of a landslides monitoring system installed in the Apennines (North Italy). Similarly, Giorgettiet al. \cite{giorgetti2016robust} deployed a network of 15 wireless sensors on a landslide in Torgio vannetto (Italy) observing high level of robustness in term of self-organisation, node failures, and energy consumption.

\subsection{5G and Device to Device Communication}
\label{subsec:2}

The upcoming 5G systems are envisioned to have the crucial capabilities such as network flexibility, (re)configuration and resilience and therefore, expected to play a key role in improving disaster situation communications. Furthermore, in 5G, network will provide media independent handover (IEEE 802.21 support) allowing seamless hand-off between various available networks thereby enabling disaster communication without any disruption. 5G networks are not only expected to attain much faster transmission throughput, but also support the emerging use-cases related to the IoT, Machine Type Communications (MTC), broadcast-like services and lifeline communications during natural disasters. 5G will fulfill these demands by adopting new technologies like proximity services, through which devices communicate with each other directly instead of relying on base stations (eNodeB) of network operators \cite{rawat2015towards}.

Device-to-Device (D2D) communication has also been used in disaster scenarios (e.g. for public safety and warning messages) to manage the radio spectrum and energy consumption for providing high Quality of Experience (QoE) and better Quality of Service (QoS). In disaster, the effective use of the radio resources is of extreme importance with the goal of serving a large number of affected people to collect information from different nodes in the disaster zone. In this context, D2D communication will be an effective solution allowing an efficient spectrum allocation without adding any further delay in content uploading for the User Equipments (UEs) \cite{rawat2015towards}.

\subsection{Software Defined Radio}
\label{subsec:2}

While LTE provides a solution to address the lack of broadband connectivity in disaster network, Software Defined Radio (SDR) technology provides a solution to address the lack of interoperability in a wireless communication scenario e.g., in military applications. SDR enables a platform to interface and communicate with different communication technologies. SDR technology could be used to support various wireless communications technologies on the same radio platform. It is also essential to define a common waveform to support the wireless backbone network. Though SDR is a promising technology, its potential application in the disaster management requires addressing various issues, for example: (1) Military oriented solutions for SDR equipment are rather costly for disaster applications, (2) Waveform processing in SDR need significant energy and computing resources that is a problem for handheld terminals.

\subsection{Cognitive Radio (CR)}
\label{subsec:2}
Public safety agencies are increasingly using wireless communication technologies to monitor disaster conditions using video surveillance cameras and sensors. The increasing use has led to congestion \cite{doumi2006spectrum} in radio frequency channels allocated to the agency. In order to address the aforementioned issue of optimum resource allocation during emergency response, CR technology could be exploited to replace the current state-of-the-art channel allocation protocols with an adaptive CE \cite{gorcin2008public}.

\subsection{Indoor Position Technologies}
\label{subsec:2}

In disaster scenarios, Global Navigation Satellite Systems (GNSS) based positioning is used to enhance the coordination of the rescue teams. However, due to the lack of GNSS coverage in indoor environments (such as tunnels and buildings), indoor navigation is required for providing the location services to first time responders. In order to make indoor positioning a potential technology for disaster scenario, some of the issues that need to be addressed are: (1) indoor-positioning devices should not be cumbersome to enable their easy deployment, (2) designing energy efficient algorithms for indoor positioning to maximize the battery lifetime of the mobile nodes in a localization system for disaster scenario \cite{robles2010low}.

\subsection{Disaster Situation Aware Protocols for Mobile Devices}
\label{subsec:2}
The integration of context-aware computing with mobile devices enable them to adapt and react to dynamic changes in the environment. This concept is used in \cite{ramesh2014context} to design a context-aware ad hoc network for effective crowd disaster mitigation by issuing an alert to prevent a stampede in the crowded area. The authors designed Disaster Aware Protocols (DAP) taking into account the disaster situation, allowing mobile devices to be effective in a disaster scenario. In the absence (or partial presence) of an infrastructure, a mobile device should be able to operate in a disaster mode serving as a lifeline for the common people on the ground. DAP for mobile devices should feature communication mode switching scheme in which information such as amount of remaining battery, mobility (mobile phone’s movement), and number of neighboring mobile devices could used by the mobile phone to decide the apt communication mode \cite{nishiyama2014relay}.

\subsection{Mobile Phone Disaster Mode}
\label{subsec:2}
Mobile phone is a potential device in the event of a disaster scenario to be able to help us connect with family and friends, locate resources, navigate to a safer location and help others. In addition, smartphones can use their integrated sensors to help allocate scarce resources to the most affected people by collecting data to enable disaster relief teams to comprehend the unraveling situation in the disaster zone. However, due to the challenges of energy-management and connectivity, currently available smart phones are not well equipped to operate efficiently during disaster. Often, disaster victims are left helpless with poor or no connectivity.

\section{Existing IoT Standards and Future Research Directions}
\label{sec:2}

According to ITU, IoT is defined as: A global infrastructure for the information society enabling advanced services by interconnecting things based on, existing and evolving, interoperable information and communication technologies.' \cite{rose2015internet}. Maximizing the communication of hardware objects and converting the harvested data into a meaningful information without any human involvement, are the two major objectives of IoT. IoT is the combination of three basic elements: hardware, middleware, and presentation \cite{kaur2017energy}. Hardware is further divided into embedded sensors, actuators, and communication systems. The embedded sensors collect data from the monitoring area and send it to the middleware element. Middleware element processes a huge amount of received data and extracts interpretable information with the help of different data analysis tools. Visualization of processed data in an easily readable form gets transformed through the presentation element. Presentation element also processes user queries to the middleware element for necessary actions. Fig. 1 shows the block diagram of an IoT system, where different communication standards have been used (in the literature) to communicate between blocks. We will discuss two major standards: LoRa and 4G LTE.

            \begin{figure*}
            	\centering
            	\includegraphics[trim=0cm 0cm 0cm 0cm, clip=true, width=0.95\textwidth]{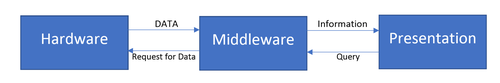}
            	\caption{IoT's three main elements and their communication }
            	\label{fig:Picture2}
            \end{figure*}

\subsection{LoRa}
\label{subsec:2}

In 2012, Semtech acquired a spread spectrum technique named LoRa. LoRa can be formed by taking the derivative of Chirp Spread Spectrum (CSS). Any MAC layer could be used with LoRa physical layer. However, the currently proposed MAC is Low Range Wide Area Network(LoRaWAN) which works on the principle of simple star topology. LoRa supports a star topology; therefore, it can transmit over a very long distance. Gateway, which is connected to a backbone infrastructure, is directly connected with the nodes. These gateways are powerful devices capable of receiving and decoding a number of concurrent transmissions (up to 50) through powerful radios. Node devices are classified into three classes: (1) Class A end devices: Transmission from node to the gateway only occurs when needed. After transmission receive window is activated to obtain the queued messages through gateway (2) Class B end-devices with scheduled receive slots: Class B operates on a similar principle as Class A node, but with additional receive window (3) Class C end-devices with maximal receives lots: These nodes are not suitable for battery powered operations due to continuous listening.

\subsubsection{Limitations of LoRaWAN}

The two main pillars of the IoT growth are transportation and logistics. Efficiency of multiple applications is targeted in areas such as disaster management, public and goods transportation. However, some applications are resistant to swing, delay and fluctuations, while others are not. Delay constraints are diverse for different applications but LoRaWAN being a low power wide area network (LPWAN) solution, is not well-suited for these applications. Contrarily, LoRaWAN supports solutions such as fleet management and control. Similarly, for video surveillance, MJPEG, MPEG-4, and H.264 are the most commonly used digital video formats for IP-based video systems. The data rate recommended for IP surveillance cameras ranges from 130 kb/s with low-quality MJPEG coding to 4 Mb/s for 1920x1080 resolution and 30 fps MPEG-4/H.264 coding \cite{adelantado2017understanding}. The data rate of LoRaWAN ranges from 0.3 to 50 kb/s per channel, thus it is not well-suited for these applications.

\subsection{4G LTE}
\label{subsec:2}
4G LTE is ideal for IoT application not only for its flat all-inclusive nature of IP architecture but also because it has built-in security along with robust and scalable traffic management capabilities. The spectral-efficiency of LTE is greater than second generation (2G) and third generation (3G) networks; therefore, data transmission could be done at a much lower rate. In this regard, data transmission is 2-3 times less costly than 3G while 20 times less than 2G. IoT friendly LTE chipsets are the foundation for the new wave of LTE device development, which are flexible, efficient, and low cost. Numerous LTE chipsets and modules are available today. These innovative solutions provide all required features to build a robust and long-life LTE devices for numerous applications at a low cost. Features includes a small footprint and ultra-low power consumption.

\subsubsection{Limitations of 4G LTE}
4G systems are mainly designed to deal with Human-type Communication (HTC) traffic. Consequently, when considering 5G systems, IoT dictate to simultaneously handle the presence of HTC and Machine-type Communication (MTC) traffic, while meeting the requirements of these traffic types. As a further step, the disruptive technologies are aiming at introducing flexibility, customization and re-conﬁgurability of the network in both radio and core segments, in order to enable the provisioning of enhanced IoT services. Indeed, a natural evolution of connecting devices to the Internet is to remotely control these devices through the Internet. However, for large number of users, 4G LTE suffers from high delay and high packet loss. The work presented in \cite{lloret2017architecture} revealed that the efficiency of 4G  LTE decreases dramatically as the amount of traffic increases.

\subsection{Research recommendations/future directions}
\label{sec:3}
Indeed, the next generation IoT, WSNs, 5G, and big data analytics stands as a major enabler to realise future intelligent infrastructures for enhanced disaster management. The widespread demand for data and the emergence of new services are inevitably leading to the so-called Resource Crisis. Hence, the evolution of the current centralised model of networked systems to new paradigms such as low power high data rate cognitive networks present a suitable path to counteract this crisis.

The existing LoRa provides transmission parameters such as transmission power,  coding rate, spreading  factor, and bandwidth, resulting  in  over  936 possible  combinations. These configuration parameters could be optimally tuned to acquire optimized bit-rate, airtime, and energy consumption, taking into account the local electromagnetic environment, constraints, and objectives. For example, increasing spreading factor to improve link reliability nearly halves the  datarate  and  doubles the energy  consumption. Similarly, increasing bandwidth doubles the datarate and halves the energy  consumption and airtime, reducing link reliability due to additional unwanted noise.

Recent research on LoRa/LoRaWAN has mainly focused on LoRa  performance  evaluation  in  terms  of  coverage,  capacity, scalability  and  lifetime \cite{oliveira2017long}\cite{petajajarvi2017performance}\cite{hosseinzadeh2017empirical}.  Furthermore,  recent  work  has also  proposed adaptive approaches to allocate optimal transmission  parameters \cite{bor2017lora}. However, most of these methods are based on state-of-the-art mathematical/statistical models and suffer from limited modelling assumptions, limited learning, inability to deal with non-linear complex behaviours, poor scalability, and no time-series/temporal data exploitation. Future research should focus on developing robust fair data rate allocation and
power control methods to address existing LoRa limitations and acquire optimised airtime, datarate, and energy consumption.

\section{Conclusion}
The emergence of new wireless communication services and demand for Big Data processing in real-time poses new architectural and radio resource management challenges. This chapter surveys research on emerging wireless communication technologies for effective disaster monitoring and management systems. WSN and IoT stands as a major enabler for enhanced disaster monitoring and management systems. In this chapter, limitations of two major IoT standards (LoRA and 4G LTE) are presented with some future research recommendations. It is concluded that future research should focus on developing artificial intelligence/machine learning driven more robust radio resource management strategies to enable optimised operations in real-time.
\bibliography{main.bbl}
\end{document}